\documentclass[12pt,a4paper]{article}
\usepackage{jheppub}

\usepackage{graphicx}
\usepackage{subcaption}
\usepackage{xspace}

\usepackage[utf8]{inputenc}

\usepackage[style=numeric-comp,sorting=none]{biblatex}
\abstract{Without significant changes to data organization, management, and
access (DOMA), HEP experiments will find scientific output limited by how
fast data can be accessed and digested by computational resources. In this
white paper we discuss challenges in DOMA that HEP experiments, such as
the HL-LHC, will face as well as potential ways to address them. A research
and development timeline to assess these changes is also proposed.}

\begin{document}

\noindent
\begin{tabular*}{\linewidth}{lc@{\extracolsep{\fill}}r@{\extracolsep{0pt}}}
 & & HSF-CWP-2017-04 \\
 & & April 3, 2018 \\ % use \date or hardwire e.g. December 15, 2017
 & & \\
\end{tabular*}
\vspace{2.0cm}

\title{HEP Software Foundation Community White Paper Working Group --
Data Organization, Management and Access (DOMA)}

\author[2]{Dario Berzano,}
\author[16]{Riccardo Maria Bianchi,}
\author[2]{Ian Bird,}
\author[13]{Brian Bockelman,}
\author[2]{Simone Campana,}
\author[18]{Kaushik De,}
\author[2,d]{Dirk Duellmann,}
\author[17]{Peter Elmer,}
\author[11]{Robert Gardner,}
\author[15]{Vincent Garonne,}
\author[6]{Claudio Grandi,}
\author[5,a]{Oliver Gutsche,}
\author[9,c,d]{Andrew Hanushevsky,}
\author[5,a]{Burt Holzman,}
\author[5,a,d]{Bodhitha Jayatilaka,}
\author[10]{Ivo Jimenez,}
\author[8,d]{Michel Jouvin,}
\author[2,d]{Oliver Keeble,}
\author[1]{Alexei Klimentov,}
\author[4]{Valentin Kuznetsov,}
\author[1]{Eric Lancon,}
\author[2]{Mario Lassnig,}
\author[3]{Miron Livny,}
\author[10]{Carlos Maltzahn,}
\author[12,b]{Shawn McKee,}
\author[7]{Dario Menasce,}
\author[5,a]{Andrew Norman,}
\author[17]{Jim Pivarski,}
\author[11]{Benedikt Riedel,}
\author[2]{Markus Schulz,}
\author[14]{Horst Severini,}
\author[10]{Michael Sevilla,}
\author[5,a]{Elizabeth Sexton-Kennedy,}
\author[5,a]{Eric Vaandering,}
\author[11]{Ilija Vukotic,}
\author[10]{Noah Watkins,}
\author[1]{Torre Wenaus}
\author[10]{and Frank Wuerthwein}
\affiliation[1]{Physics Department, Brookhaven National Laboratory, Upton, NY, USA}
\affiliation[2]{CERN, Geneva, Switzerland}
\affiliation[3]{Center for High Throughput  Computing, Computer Sciences Department, University of Wisconsin-Madison, Madison, WI, USA}
\affiliation[4]{Cornell University, Ithaca, NY, USA}
\affiliation[5]{Fermi National Accelerator Laboratory, Batavia, IL, USA}
\affiliation[6]{INFN Sezione di Bologna, Università di Bologna, Bologna, Italy}
\affiliation[7]{INFN Sezione di Milano-Bicocca, Milano, Italy}
\affiliation[8]{LAL, Université Paris-Sud and CNRS/IN2P3, Orsay, France}
\affiliation[9]{SLAC National Accelerator Laboratory, Menlo Park, CA, USA}
\affiliation[10]{University of California, San Diego, La Jolla, CA, USA}
\affiliation[11]{Enrico Fermi Institute, University of Chicago, Chicago, IL, USA}
\affiliation[12]{Department of Physics, The University of Michigan, Ann Arbor, MI, USA}
\affiliation[13]{University of Nebraska-Lincoln, Lincoln, NE, USA}
\affiliation[14]{Homer L. Dodge Department of Physics and Astronomy, University of Oklahoma, Norman, OK, USA}
\affiliation[15]{Department of Physics, University of Oslo, Oslo, Norway}
\affiliation[16]{Department of Physics and Astronomy, University of Pittsburgh, Pittsburgh, PA, USA}
\affiliation[17]{Princeton University, Princeton, NJ, USA}
\affiliation[18]{Department of Physics, The University of Texas at Arlington, Arlington, TX, USA}
\note[a]{Supported by the US-DOE, DE-AC02-07CH11359}
\note[b]{Supported by the US-DOE, DE-SC0007859 and US-NSF, 76749/1136652/2}
\note[c]{Supported by the US-DOE, DE-AC02-76SF0051}
\note[d]{CWP Data Organisation, Management and Access paper Editorial Board member}

\maketitle

\newpage

\section{Introduction}\label{introduction}

The coming decade will bring HEP to the exabyte scale with the expected data
volumes of the HL-LHC~\cite{HL-LHC,1742-6596-515-1-012012,Apollinari:2284929} experiments, the DUNE~\cite{Acciarri:2015uup} experiment, and even non-HEP
experiments such as the Square Kilometre Array (SKA)~\cite{SKA} entering this
regime. In devising computing models for this era, many
factors have to be taken into account. In particular, the increasing
availability of very high-speed networks may reduce the need for
CPU and data co-location. Such networks may allow
for more extensive use of data access over the wide-area network (WAN) while providing fail-over capabilities, global and federated data
namespaces, and will have an impact on data caching. Shifts in the data
presentation and analysis models, such as a potential move to event-based or
sub-event-based data streaming from the more traditional dataset-based or file-based
data access, will be particularly important for optimizing the
utilization of opportunistic computing cycles on HPC facilities,
commercial cloud resources, and campus clusters, and can potentially
resolve currently limiting factors such as job eviction.

Planned analysis of HL-LHC data and data analysis from other
international collaborations will need to adopt a distributed computing
model due to their scope. The data management systems that will enable
this type of analysis will need to provide cost-based optimization of
the data locations and delivery mechanisms.

\section{Challenges and
Opportunities}\label{challenges-and-opportunities}

The LHC experiments currently collectively provision and manage about an Exabyte of
storage of which approximately half is archival and half is
traditional disk storage. The annual storage requirements are expected
to jump by a factor of 10 for the HL-LHC. This increase is faster than
projected density gains in current storage technologies and thus will present major challenges. Storage
will remain one of the visible cost drivers for HEP computing. The projected growth and cost of the computational resources needed to
analyze the data is expected to grow even faster than the base
storage costs. The combination of storage and analysis computing costs
may restrict scientific output and potential physics reach of the
experiments. Thus, new techniques and algorithms are likely to be
required.

The three main challenges for data in the HL-LHC era can be
summarized as:

\begin{enumerate}
\def\labelenumi{\arabic{enumi}.}
\item \textbf{Big Data:} The expected data volume will significantly increase in the HL-LHC era. Computing systems will need to
  handle this without significant cost increases and within evolving
  storage technology limitations.
\item \textbf{Dynamic Distributed Computing:} In addition, the significantly
  increased computational requirements for the HL-LHC era will also
  place new requirements on data. Specifically, the use of new types of
  compute resources (e.g., cloud, HPC) with different dynamic %don't want this to be just cloud and HPC, hence "new types"
  availability and characteristics will require more dynamic
  DOMA systems.
\item \textbf{New Applications:} New applications, such as machine learning
  training or high-rate data-query systems for analysis, will likely be
  employed to meet the computational constraints and to extend the
  physics reach of the HL-LHC. These new applications will place new
  requirements on how and where data is accessed and produced. For
  example, specific applications (e.g., training for machine learning)
  may require use of specialized processor resources, such as GPUs.
\end{enumerate}

The rapid increase in recent years of data-intensive problems in both
the commercial world and in the rest of the research world also provides
a number of opportunities and solutions to tackle these challenges.

\section{Current Approaches}\label{current-approaches}

The original LHC computing models (developed circa 2005) were built up from
simpler models used before distributed computing was a central part of
HEP computing. This allowed for a reasonably clean separation between
three different aspects of interacting with data: organization,
management and access.

\begin{itemize}
\item \textbf{Data Organization:} This is essentially how data is structured
  as it is written. Most data is written in the ROOT~\cite{Brun1996} file format. ROOT serializes data objects, compresses, and writes them in a column-wise manner. The data written in this way can only be meaningfully read by a software that includes data object libraries.
\item \textbf{Data Management:} The key challenge here was the transition to
  the use of distributed computing in the form of the grid. To tackle this, the
  experiments developed dedicated data catalog,  transfer and placement systems. To first
  order, the computing models were rather static: data was placed at
  sites and the relevant compute jobs were sent to the right locations.
\item \textbf{Data Access:} Various protocols are used for direct reads
  (RFIO, dCap, XRootD, etc.) within a given computer center and/or
  explicit local stage-in and caching for read by jobs. Application
  access may use different protocols than those used by the data
  transfers between sites.
\end{itemize}

Before the LHC turn-on and in the first years of data taking, these three
areas were, to first order, optimized independently. Many of the
challenges were in the area of ``Data Management'' (DM) as the Worldwide
LHC Computing Grid~\cite{WLCG} was commissioned. As the LHC experiments' computing matured
through Run1 and Run2, interest has turned to optimizations spanning
these three areas. For example, the recent use of ``Data Federations''
must consider together both Data Management and Access. As we will see below, some of the
foreseen opportunities towards HL-LHC may require global optimizations.

Thus, in this document we take a broader view than traditional DM,
and consider the combination of ``Data Organization, Management and
Access '' together. We believe that this full picture of data
needs in HEP will provide important opportunities for efficiency and
scalability as we enter the many-Exabyte era.

\subsection{HEP Workflow in Context}\label{hep-workflow-in-context}

HEP data transfer and access patterns require network and storage architectures to be able to sustain a large dynamic range in IOPS. The primary data analysis workflows that drive
these bandwidth requirements can be categorized into four main
activities:

\begin{enumerate}
\def\labelenumi{\arabic{enumi})}
\item \textbf{Reconstruction:} Event reconstruction is CPU limited, due to
  the complexity of detector data and the computational algorithms
  needed to associate and disentangle the data. This is true of
  reconstruction methods used today at the LHC and throughout the HEP
  community and will continue to be the driving characteristic of
  reconstruction at the HL-LHC and for liquid argon TPC experiments. The
  CPU-bound nature of the algorithms and the ratio of event size to
  network bandwidth make the option of streaming individual events
  records (e.g., full collision events or beam spill triggers)
  across the WAN to compute elements a feasible
  strategy. The requirements of event streaming can be
  satisfied by a wide range of storage solutions.
\item \textbf{Simulation:} HEP event simulation strategies focus on
  transforming a small set of key input parameters into event and
  detector response information that is representative of what would be
  seen in the physical detector systems. Simulation processes are
  typically characterized by highly asymmetric IO. They typically have very small data ingest (typically just the key input
  parameters), while output data are similar in size and
  structure to real detector data. Due to their similar characteristics, output data from simulation processes share the requirements and associated challenges of detector data. The primary drivers that differentiate
  this type of HEP workflow from the other categorizations are the
  increased complexities that the simulation stages often require. These
  simulations often rely on auxiliary event generation mechanisms and
  can, in some HEP domains, require large external data sets
  representing interaction cross sections or detector response functions
  to be available to each instance of the event simulation. This type of
  common input and overlay data may present a challenge in relation to
  its scaling to the runtime environments of future HPC facilities and
  to the design of caching layers at computing site.
\item \textbf{Analysis:} Analysis level datasets and subsets are typically
  comprised of refined or reduced information that is most pertinent to
  the extraction of physics quantities. These analysis-focused event
  records can often be accepted or rejected through fast pre-filtered
  methods which do not require the reading and retrieval of the full
  event record. This fundamental difference in the way that the analysis
  data is consumed makes it more susceptible to storage bandwidth
  limitations and to IOPS transaction limits that underlying
  technologies can have. In particular, the internal structure of the
  event data can dramatically impact the efficiency of data retrieval
  and favor different access models and protocols while the actual analysis calculations being performed can
  dramatically affect the CPU-to-IO ratios. These factors make this
  categorization of data the most challenging to project to future
  storage needs and models.
\item \textbf{Replication:} Data replication, whether within a site or
  across multiple sites, is performed to enable efficient data access by
  improving data locality (the proximity of the data to compute
  resources) and exploiting the available data bandwidth between the
  storage and the compute locations. Data replication also serves to
  guard against data loss. Data replication places modest requirements
  on the IOPS that need to be supported, but drives
  the bandwidth requirements that are needed in enacting site-to-site
  data transfers between WAN endpoints (e.g., 10-100~Gbps). This
  requirement on the available WAN bandwidth provides a corresponding
  performance requirement on on the underlying storage systems to match
  network performance. Satisfying both requirements is necessary to
  provide efficient point-to-point data flow.
\end{enumerate}

In addition to the characteristics described above, HEP workflows in
each category exhibit a high degree of parallelism, which allow the work
to map readily into both high throughput computing (HTC) and HPC environments. These domains span wide
dynamic ranges in their requirements for concurrent data access. If the
experiments of the HL-LHC, DUNE and other international projects
are to utilize these facilities, they will need the ability to integrate
with multiple storage technologies. The storage technologies will need
to be tuned or matched to the corresponding computing resources that the
experiments are targeting.

These storage models will also need to be designed and optimized for
cost, in the context of the scientific workflows that they are enabling.
Simultaneously, these storage models must remain flexible enough to
support evolving physical storage options similar to the the shifts that
have occurred in recent years between NVMe, SSDs, shingled disks, tapes,
and other technologies.

\section{Data Organization Models}\label{data-organization-models}

The HEP community has long-established data organization patterns that have
shaped the prevailing data access and storage
models that are in use today. Today's HEP data organization is file-centric with sets of events stored in a given file. Alternative data organization models could yield gains depending on their respective data access and storage models. These
alternative designs should be evaluated for the HL-LHC era, as they may produce cost reduction, better network utilization, and efficiency gains.

There are two main avenues that these organizational changes can follow,
each with different impacts on the analysis methodologies. First, we
should reconsider the structure of the data, which can be row-oriented,
column-oriented, have varying levels of granularity, or represent larger or
smaller aggregations of objects, boosting the efficiency of
particular applications.

Secondly, we should reconsider the granularity at which data management
systems operate. This is currently at aggregations of events (files, or
larger sets, such as blocks or datasets), and could shift towards being
the event (or even sub-event), potentially reducing storage volumes. In
that case, other types of ``containers'' beyond files could be considered
(e.g. objects).

Design decisions will be subject to competing requirements from data
management and access optimization, and will have consequences
throughout the system. We have to tackle a global optimization problem,
taking into account all use cases and their respective frequencies to
further optimize application performance for a given resource
investment.

\subsection{Optimizing Data Access}\label{optimizing-data-access}

\subsubsection{Data structures}\label{data-structures}

The structure of data has to be adapted to application requirements.
At one limit, the data could be organized simply as events, while the other
extreme is to organize the data as columns which represent sub-event
level objects such as particles. Event-based organization favors
applications like event displays while column-based organization favors
large statistical analysis. Our current system of data tiers in files
draws intermediate (and overlapping) rectangles between these two
extremes. Finding the right balance will be an important factor in
efficient data analysis.

While it appears that organizing data in a more fine-grained fashion, as
events or columns, has the potential of opening additional avenues for
cost-savings and flexibility, the implications of such an organizational
shift are still not well understood and need to be investigated.

\subsubsection{Object and Cloud Storage
systems}\label{object-and-cloud-storage-systems}

Decisions on granularity and structure have the potential of broadening
the scope of technologies that can be used to store the data; for
example, exploiting object stores (such as AWS S3 or Ceph~\cite{Weil:2006:CSH:1298455.1298485}) as event stores. Such an approach
would support parallel reading and writing of events as
event objects from applications. This could allow applications to better
scale to large numbers of CPU cores whereas in current models application
concurrency is limited by the requirement to sequentially write events
to files. Currently, the efficiency of of reading and writing events
from object stores is sensitive to the size of the event---
the smaller the event, the worse the efficiency is. Some of these potential inefficiencies
may be mitigated by creating event bundles (i.e. several events in one object). The
evolution of the ATLAS Event Service may provide some of the answers and
should be supported. As a parallel effort, the ROOT team is also looking
into increasing concurrency within the ROOT framework. This effort
should also be supported as it may have more immediate success in the
near term.

Commercial providers offer innovative storage services (``cloud
storage'') which may fit certain use cases well. The relevance of these
solutions to HEP computing models must be understood, not only through
technical metrics such as performance, but with an understanding of the
underlying cost structure and the risks associated with such
procurements. Reliance on commercially procured cloud storage as a core
component of data organizational models would require a potential
fundamental shift in user policies and discussion with funding agencies.

\subsection{Optimizing Data Management}\label{optimizing-data-management}

Data Management policy will directly affect data access. For example,
the decision to distribute multiple enriched data-samples optimized for
different use cases would improve data access and thus compute
efficiency, at the expense of larger volumes transferred and stored.
Similarly, the unit of data management could be changed, and become the
event, which promises to reduce storage volumes and promote the use of
opportunistic resources through enabling pre-emptibility in jobs. This
would have cascading effects on data catalogs, data distribution and
storage systems, and would affect data access. Where an experiment
positions itself amongst all these possibilities will be dictated by its
policy. Thus the potential gains in terms of efficient use of resources
should be realized through data management systems which allow this
policy to be expressed and implemented.

\subsubsection{Data catalog models}\label{data-catalog-models}

A key requirement in devising data catalog models is understanding the
required granularity of data to be cataloged and what data for an
experiment needs to be centrally cataloged or tracked, and for what
purpose (accounting, metadata, location). Analysis access to the data,
for instance, may benefit from being below the event level (e.g. physics
object). Any such catalogs may have to be external to the primary data
store (e.g., relational databases that work with metadata in the primary
data store). If cataloging requirements for production and
reconstruction differ considerably from analysis, an approach with
multiple, complementary catalogs can also be considered. The cataloging
schemes chosen will be tightly coupled with the data organization
models. For example, in the case where data organization is object-based
or content-addressable storage is used, the catalogs need only a single
handle to have a fully descriptive location of a given event or other
object being sought.

Our concern is that the size and complexity of a data catalog is
proportional to the granularity of the data organization. Given that the
organization is still in a state of flux (see previous section) it is
difficult to judge the impact of future dataset sizes on data catalog
models other than the simple fact that catalog sizes are likely to grow.
It does seem clear, however, that data catalog models need to evolve in
lockstep with the evolution of data organization models. A large and
diverse set of foundational technologies exist upon which such future
catalogs could be built. An informed technology choice which supports
the anticipated catalog use cases will be crucial in enabling future
scalability.

~

\subsubsection{Data delivery methods}\label{data-delivery-methods}

Data delivery models can become a limiting factor in performance,
especially as they heavily depend on shared resources such storage
systems and the network. Caching has to be considered to avoid data
delivery bottlenecks and unnecessary transfers. At present, HEP data
delivery models are designed solely with the traditional sequential
processing paradigm used in reconstruction in mind. Computing models
where the only form of persistent storage is on tape and all, or nearly
all, disk-based storage is considered as a caching layer should also be
considered, in which case work on optimizing workflows for cache
efficiency and exploiting data popularity metrics would be valuable.

~

A useful analogy for data delivery methods might be content delivery
network services (CDNs) such as the one offered by Akamai. Without CDNs,
on-demand streaming of movies over the Internet or even the delivery of
the web sites of major news outlets like CNN or BBC would quickly
collapse. What is the equivalent of content delivery for accessing huge
datasets over distance where a particular analysis might only
require a fraction of the dataset? CDNs use, to great benefit, highly
sophisticated proprietary naming, caching, and placement techniques
that are customized to particular types of data and what is known about
their access patterns. Similarly intelligent techniques could be used to
place, compose, and deliver scientific data products, minimizing data
movement, latency, and other costs, taking advantage of data semantics
and information about the location of reusable intermediate results.
This is not unlike query planning in relational databases, except for
the scale and the different data type of scientific data: a query
optimizer might be able to recognize that data can be filtered remotely,
or combined from data products cached in the neighborhood instead of
having to fetch everything across continents. New approaches, like Named
Data Networking (NDN), could play a role. NDNs allow the naming, at the
network level, of the data you are looking for rather than its location, which
opens the paths for many optimizations behind the scenes based on
factors such as the client application location or the data type.

Currently, new data-caching strategies are being deployed (e.g., XRootD
caching proxy servers, server-less caching, ARC-style caching) and may
provide sufficient information on how such strategies can optimize
mammoth data delivery task for HL-LHC event reconstruction and analysis.
However, without a global view on how these strategies are used, they
will likely have a limited impact on improving data delivery efficiency.

\section{New Analysis Paradigms}\label{new-analysis-paradigms}

Data analysis in most of HEP is currently tied to ROOT-based formats. In many
currently-used paradigms, physicists consider all events at an
equivalent level of detail and in the format offering the highest level
of detail that needs to be considered in an analysis. ~However, not
every event considered in analysis requires the same level of detail.
One consideration to improve event access throughput is to design event
tiers with different abstractions, and thus data sizes. All events can
be considered at a lighter-weight tier while events of interest only can
be accessed with a more information-rich tier.

For more scalable analysis, another opportunity to evaluate is how much
work can be offloaded to a storage system, for example caching
uncompressed or reordered data for fast access. The idea can be extended
to virtual data and to query interfaces which would perform some of the
transformation logic currently executed on CPU workers. Interactive
querying of large datasets is an active field in the Big Data industry;
examples include Spark-SQL, Impala, Kudu, Hawq, Apache Drill, and Google
Dremel/BigQuery. A key question is about the usability of these
techniques in HEP and we need to assess if our data transformations are
not too complex for the SQL-based query languages used by these
products. We also need to take into account that the adoption of these
techniques, if they prove to be beneficial, would represent a disruptive
change which directly impacts the end user and therefore promoting
acceptance through intermediate solutions would be desirable. One such solution - ``Service X'' - has been proposed in \href{https://bit.ly/2Ix9NIL}{Organizing Data Lakes}. The service would start with the data in its persistent format, perform one or more operations on it  (e.g. joins, query, decompression, filtering to sub-event level), and finally deliver a data stream to a client.

Many analyses may benefit from column-based data access instead of the
more traditional row-based access. Enabling data queries that consider
histogram indexing is another feature that could provide performance
increases in analysis.

When evaluating the potential benefits of moving to new techniques, for
analysis in particular but not only, we should not forget that some
techniques, like machine learning-based techniques, may have
contradictory requirements. On one end, machine learning should
dramatically reduce the pressure on the storage at the cost of using
more CPU but it requires a learning phase which, even if it is shorter
compared to the exploitation phase, can require a significant amount of
resources with a non-typical, iterative, data access pattern. Thus, the
challenge remains to find a good compromise to efficiently support a
large variety of access patterns.

In addition to assessing the techniques that could improve the
performance for the various use cases that we have, it is also important
to ensure that the resources required by these techniques are in line
with expected budget scenarios. Many of these techniques may imply
shifting some of the storage resource usage to CPU usage or vice-versa.
We need to ensure that the global cost will remain similar to what it is
today and that the resulting computing model evolution is compatible
with other constraints.

\section{Performance: Tools and
Metrics}\label{performance-tools-and-metrics}

As the size of available datasets grows, we must allow for the
possibility of increased user-driven selection and dataset generation.
In terms of reducing the strain on data access mechanisms, techniques such as data
augmentation can prove useful. Cluster-computing frameworks (e.g.,
Spark) can also be leveraged to reduce the time of dataset reduction and
access. It will be necessary to study and classify I/O patterns in
applications used in order to understand how data access methods can be
optimized across prevalent access patterns. Pattern data can then be
used in algorithms resulting from active research in computer science
that studies trade-offs between storage and CPU in various computing models.

A first correlated analysis of the I/O patterns of different
computational tasks has already started at the CERN Data Centre and
within several experiments. In particular, the combination of
infrastructure and experiment information (e.g. detailed knowledge of
hardware capabilities and network topology, per job CPU and storage
metrics) may allow us to define more meaningful performance metrics than
the traditionally-used raw data rate and CPU utilization which both have
their limitations. These metrics should enable a joint quantitative
comparison of different storage and computing approaches across
individual system boundaries and thus lead to a more effective resource
investment.

~

\section{Common Challenges}\label{common-challenges}

The projected event complexity of data from future LHC runs and from
high resolution liquid argon detectors will require advanced
reconstruction algorithms and analysis tools to understand. The
precursors of these tools, in the form of new machine learning paradigms
and pattern recognition algorithms, already are proving to be drivers
for the CPU needs of the HEP community. As these techniques continue to
gain traction in our field and evolve, they will place new requirements on the computational
resources that need to be leveraged by all of HEP. The storage systems
that are developed and the data management techniques that are employed
will need to directly support this wide range of computational
facilities, and will need to be matched to the changes in the
computational work so as not to impede the improvements that they are
bringing.

Storage will remain one of the visible cost drivers for HEP computing,
but the projected growth and cost of the computational resources needed
to analyze the data are expected to grow faster than the base storage
costs. The combination of storage and analysis computing costs may
restrict scientific output and potential physics reach of the
experiments. There must be R\&D efforts in data management on how to
minimize the impact of the data access and storage model on the overall
cost of doing scientific analysis. This R\&D should include an
optimization of both the capital costs of storage as well as the
potential impacts the storage systems can have on the CPU requirements
for the experiments and their costs.

The ability to leverage new storage technologies, as they become
available, into existing data delivery models is a challenge that we must
be prepared for. New storage systems will present new interfaces and new
behavior. A key decision in their successful exploitation will be
whether to encapsulate this within a more familiar service or whether to
present the new interface directly to applications which will thus have
to be adapted. As discussed in the preceding sections, much of this
change can be aided by active R\&D into our own IO patterns; an approach
which has not yet been adopted widely by the field.

HEP experiments should be prepared to leverage ``tactical storage''.
Such storage may be provisioned only when it becomes cost-effective
(e.g.,
from a cloud provider) and have a data management and provisioning
system that can exploit such resources on short notice. Volatile data
sources would impact many aspects of the system; catalogs, job
brokering, monitoring/alerting, accounting, the applications themselves.

On the hardware side, R\&D is needed in alternative approaches to data
archiving to determine the possible cost/performance trade-offs.
Currently, tape is extensively used to hold data that cannot economically be made available online. While the tape-stored data is still accessible,
it comes with a high latency penalty and thus limiting possible analysis. We suggest investigating either separate direct-access-based archives (e.g.
disk or optical) or new models that overlay online direct-access volumes
with archive space. This is especially relevant when access latency is
proportional to storage density. Either approach would need to also
evaluate reliability risks and the effort needed to provide data
stability.

Cost reductions in maintenance and operation of the storage
infrastructure can be realized through convergence of the major
experiments and resource providers on shared solutions. This does not
necessarily mean promoting a monoculture, as different solutions will be
adapted to certain major classes of use-case, type of site or funding
environment. Indeed, there will always be a judgment to make on the
desirability of using a variety of specialized systems, or abstracting
the commonalities through a more limited but common interface (the SRM
story illustrates this point). Reduced costs and improved sustainability
will be further promoted by extending these concepts of convergence
beyond HEP and into the other large-scale scientific endeavors that
will share the infrastructure in the coming decade. Efforts must be made
as early as possible, during the formative design phases of such
projects, to create the necessary links.

Finally, any and all changes undertaken must not make the ease of access
to data any worse than it is under current computing models. We must
also be prepared to accept the fact that the best possible solution may
require significant changes in the way data is handled and analyzed.
What is clear is that what is being done today will not scale to the
needs of HL-LHC experiments.

\section{Research and Development Roadmap and
Goals}\label{research-and-development-roadmap-and-goals}

\subsection{Sub-file granularity}\label{sub-file-granularity}
Sub-file granularity ({\it e.g.} sub-event-based or event-based granularity) should be studied
to see whether it can be implemented efficiently and in a scalable, cost-effective
manner for all use cases that rely on event selection and whether it offers an advantage
over the current paradigm of file-based granularity. Proposed actions are:

\begin{itemize}
\item Quantify impact on performance
  and resource utilization (storage, network) for the main type of
  access patterns ({\it i.e.} simulation, reconstruction, analysis).
\item Assess the impact on catalogs and
  data distribution.
\item Assess whether
  event-granularity makes sense in object stores that tend to require
  large chunks of data for efficiency.
\item Test for improvement in recoverability
  from preemption, in particular when using cloud spot resources and/or
  dynamic HPC resources.
\end{itemize}

\emph{The above tasks should be completed by 2020 (and can be performed in parallel).}

\subsection{Data organization and analysis
technologies}\label{data-organization-and-analysis}
Data organization and analysis technologies in use by other big data users should be studied
to glean whether we can benefit from them. Proposed actions are:

\begin{itemize}
\item Evaluate row-based versus column-based
  data organization on the performance of each type of access used.
\item Investigate data storage and access solutions that support the use of map-reduce or
Spark-like analysis tools and their adaptability to HEP analysis needs.
\item Evaluate new compression schemes, just-in-time
  decompression schemes and mappings onto hardware architectures
  considering the flow of data from spinning disk to memory and
  application.
\item Evaluate possible alternative storage formats to the ROOT format, especially storage formats optimized for storage efficiency and/or  data filtering through metadata
\end{itemize}

\emph{The proposed proof-of-concept involves the above tasks that should be completed by 2020.
If successful, implementation as a production system will be done in the following years.}

\subsection{Data caching}\label{data-caching}
Investigate the role that data placement optimizations, such as caching, can play in optimizing
the use of computing resources as well as the technologies that can be used for this. Proposed actions are:

\begin{itemize}
\item Quantify the benefit of caching for reconstruction, analysis, and simulation.
\item Assess the benefit of caching for
  Machine Learning-based applications, in particular for the learning
  phase.
\item Evaluate the benefits of using different approaches to data delivery from what is currently in use in HEP.
Namely, Content Delivery Networks (CDN) and Named Data Networking (NDN) should be studied.
\end{itemize}

\emph{The first two actions should be completed by 2020 but evaluation of CDN/NDN is more long-term}

\subsection{Getting the most out of the
storage diversity}\label{storage-diversity}

The landscape of increasing diversity in storage service offerings and technologies should be studied and
exploited in order to reduce HEP infrastructure costs. The proof-of-concept phase involves the following tasks:

\begin{itemize}
\item Understand what role tactical or opportunistic (potentially short-term) storage can play in HEP.
\item Re-evaluate the role of archival storage solutions for HEP.
\item Evaluate the role of "newer" storage architectures, such as object stores or key-value stores
\end{itemize}

\emph{The  proof-of-concept should be completed by 2020 and, if successful, implementation in
production will be done in the following years.}

\subsection{Global optimization of efficiency and latency}\label{efficiency-latency}
The inherent trade-offs between data access latency and efficiency of CPU use need to be studied and optimized
on a global scale across the HEP infrastructure. The proof-of-concept phase involves the following tasks:

\begin{itemize}
\item Understand the impact of concentrating the
  data in fewer, larger locations (the ``data-lake'' approach).
\item Understand the impact of an increased use of
  opportunistic compute resources, specifically those further from where data are stored.
\end{itemize}

\emph{The  proof-of-concept should be completed by 2020 and, if successful, implementation in
production will be done in the following years.}

\subsection{Data Access and Data Access Patterns}\label{access-patterns}
The data access needs and patterns should inform the the choice of storage media and data location. Proposed areas of inquiry are:

\begin{itemize}
\item Understand data access patterns, including investigating frequency of access of data
\item Exploring ways to ease preliminary data filtering steps
\item Exploring commercially-popular data transfer methods (e.g. HTTP)
\end{itemize}
\emph{The  proof-of-concept should be completed by 2020 and, if successful, implementation in
production will be done in the following years.}

\section{Conclusions}\label{conclusions}

This document presents several
areas pertaining to data access and management in HEP that will need to
be re-examined in the coming decade before the expected volume and
complexity of data becomes prohibitively expensive to store, access, and
analyze. Extending current data handling methods and methodologies will
prove intractable in the HL-LHC and DUNE era. The development and
adoption of new data analysis paradigms gives the field, as a whole, a
window in which to adapt our data access and data management schemes to
ones which are more suited and optimally matched to a wide range of
advanced computing models and analysis applications. This type of shift
has the potential for enabling new analysis methods and allowing for an
increase in scientific output.

\sloppy
\raggedright
\clearpage
\printbibliography[title={References},heading=bibintoc]

\end{document}